\documentstyle[amssymb,aps,prb,twocolumn,psfig]{revtex}

\begin{document}

\wideabs{
\title{Superconducting current-phase relation in Nb/Au/(001)YBa$_2$Cu$_3$O$_x$ heterojunctions}

\author{P. V. Komissinski$^{1,3}$, E. Il'ichev$^2$, G. A. Ovsyannikov$^3$,
  S.A. Kovtonyuk$^3$, M. Grajcar$^4$,  Z. Ivanov$^1$,  Y. Tanaka$^5$,  N. Yoshida$^5$, and S. Kashiwaya$^6$}

\address{$^1$Department of Microelectronics and Nanoscience, Chalmers University of Technology,
and Goteborg University, S-412 96, Goteborg, Sweden}

\address{$^2$Institute for Physical High Technology, Dept. of
Cryoelectronics, P.O. Box 100239, D-07702 Jena, Germany}

\address{$^3$ Institute of Radio Engineering and Electronics RAS, Moscow
103907, Russia}

\address{$^4$ Department of Solid State Physics, Comenius University,
Mlynsk\'{a} Dolina, SK-84215 Bratislava, Slovakia}

\address{$^5$Department of Applied Physics, Nagoya University, 464-8603,
Nagoya, Japan}

\address{$^6$Electrotechnical Laboratory, Umezono, Tsukuba, Ibaraki, 
305-8568,Japan}
\date{\today }
\maketitle

\begin{abstract}
We probe directly the superconducting current-phase relations (CPR)
of Nb/Au/(001) YBa$_{2}$Cu$_{3}$O$_{x}$ heterojunctions prepared
in epitaxial $c$-axis oriented YBa$_{2}$Cu$_{3}$O$_{x}$ (YBCO) thin films.
The measurements are performed by a single-junction interference experiment.
Atomic Force Microscope observations and estimations made from $I-V$
characteristics confirm the $c$-axis current transport. No deviation of the
current-phase relation from a sinusoidal shape has been found within the
experimental accuracy. Possible reasons of the absence of the second
harmonic in the CPR due to a second order tunneling are discussed.
\end{abstract}


}%
%

The symmetry of the order parameter in high temperature superconductors has
been discussed since their discovery.\cite{Bickers} Several types of phase
sensitive experiments have been performed, such as corner-junction SQUID\cite
{Wollman93} and tricrystal ring experiment.\cite{Tsuei} The results point
upon a dominant $d$-wave symmetry of the order parameter. In contrast, the
observation of a Josephson current in $c$ direction between a heavily
twinned YBa$_{2}$Cu$_{3}$O$_{x}$ (YBCO) and a conventional $s$-wave
superconductor indicates the presence of $s$ component in YBCO. Recently
this conflict has been resolved.\cite{Kouznetsov} It has been shown that for
YBCO there is mixed $d$ and $s$ pairing with a dominant $d$-wave component.
In this a case, the theoretical approach developed in Ref.\cite{Tanaka94,zag}
is valid and one can expect a finite amplitude of the second harmonic (the
term $I_{2}$ in the Josephson current, $I_{s}=I_{1}\sin (\varphi )+I_{2}\sin
(2\varphi )$) of the current-phase relationship (CPR) for $c$-axis junctions
as well as for $45^{\circ }$ grain boundary junctions in the $ab$ plane. For
the latter, the second harmonic of the CPR has been recently observed.\cite
{Ilichev98a,Ilichev99} On the other hand, observations of microwave induced
steps in $c$-axis oriented YBCO-Pb tunnel junctions at multiples of $hf/2e$
imply first-order currents and suggest a dominant $s$-wave contribution to
the order parameter of YBCO.\cite{Kleiner}

In this paper we report precise measurements of CPR in Nb/Au/(001)YBa$_{2}$Cu%
$_{3}$O$_{x}$ heterojunctions on YBCO epitaxial thin films. A pure $%
I_{s}=I_{1}\sin (\varphi )$ dependence has been observed and the reasons for
the absence of the second harmonic in the CPR are discussed.

Epitaxial YBCO thin films with thicknesses of $150%
\mathop{\rm nm}%
$ were obtained $in\,situ$ by laser deposition on (100) LaAlO$_{3}$
substrates. The YBCO films were usually twinned in $ab$ plane. The
temperature of the superconducting transition was determined by magnetic
induction measurements as $T_{c}=88\div 90%
\mathop{\rm K}%
$. The YBCO film was $in\,situ$ covered by Au thus preventing the
degradation of the YBCO surface layer during processing and forming an
interface with low resistance, $r\equiv R_{N}S=10^{-6}%
\mathop{\rm %
\Omega}%
\cdot 
\mathop{\rm cm}%
^{2}$, where $R_{N}$ is the normal state resistance and $S$ is the junction
area. Direct contacts Nb/YBCO usually have interfaces with resistance of the
order of $10^{-2}%
\mathop{\rm %
\Omega}%
\cdot 
\mathop{\rm cm}%
^{2}$. Photolitography and low energy ion milling techniques were used to
fabricate Nb/Au/YBCO junctions, with $8\div 20%
\mathop{\rm nm}%
$ Au layer and $200%
\mathop{\rm nm}%
$ thick Nb films (Fig. 1). Details of the junction fabrication were reported
elsewhere.\cite{Komissinskii99}

The surface quality of the YBCO film is very important when the current
transport in the $c$-direction is investigated. High-resolution atomic force
microscope (AFM) studies reveal a smooth surface consisting of approximately 
$200%
\mathop{\rm nm}%
$ long islands with maximum peak-to-valley distance of $3\div 4%
\mathop{\rm nm}%
$ (Fig. 2). The contribution of $c$-axis and $ab$-plane resistance to the
total resistance $r$ can be estimated by using the following equation: 
\begin{equation}
r=r_{c}r_{ab}/(r_{ab}+r_{c}\tan \gamma ),  \label{eq:r}
\end{equation}
where $\gamma $ is the averaged slope angle of the islands on the surface,
and $r_{c}$ and $r_{ab}$ are characteristic interface resistances for
current transport along $c$-axis and in $ab$-plane, respectively. They are
determined by the Fermi velocity mismatch between Au and YBCO.\cite{Kupr91}
Even for a typical Fermi momentum anisotropy of about $3$ for YBCO, one
obtains $r_{c}$ one order of magnitude higher than $r_{ab}$ for sharp
interfaces.\cite{Komissinski} For the typical values $\gamma <2^{\circ }$ we
estimate that the tunneling area in $a-b$ plane is less than $3\%$ of the
total one. This fact indicates current flowing along the $c$-axis of YBCO.
The absence of Zero Bias Conductance Peaks (ZBCP),\cite{Wei,Tanaka95} which
is associated with Andreev bound states,\cite{Hu} gives additional evidence
for a $c$-axis transport in our junctions.\footnote{%
Theory predicts the appearance of ZBCP in the case of facets on the surface
even for (100)- oriented YBCO films.\cite{Fogelstrom,Tanuma}}

We have measured 20 single junctions with areas in the range of $10\times
10^{2}\mu $m$^{2}$ to $30\times 30\mu $m$^{2}$. The $I-V$ curve (see curve 
{\it a}, Fig. 3) shows the dependence typical of a superconducting tunnel
junction with critical current density $J_{c}=1\div 12%
\mathop{\rm A}%
/%
\mathop{\rm cm}%
^{2}$, normal resistance, $r=3\div 20\mu 
\mathop{\rm %
\Omega}%
\cdot 
\mathop{\rm cm}%
^{2}$ and $I_{c}R_{N}=10\div 90\,\mu 
\mathop{\rm V}%
$ ($I_{c}$ is the junction critical current). The $I-V$ curve for the small
voltage scale, $V\leq 0.3$m$%
\mathop{\rm V}%
$, is shown in the inset, Fig. 3. Fig. 4 presents the temperature dependence
of the critical current $I_{c}(T)$ normalized to $I_{c}$ at $4.2%
\mathop{\rm K}%
$. The differential resistance $vs.$ voltage dependence $R_{d}(V)$ (see
curve {\it b}, Fig. 3) exhibits a gap-like structure at $V\approx 1.2$m$%
\mathop{\rm V}%
$, corresponding to the Nb superconducting energy gap, $\Delta _{Nb}$. This
structure has a BCS-like temperature dependence and disappears at $T\approx
9.1%
\mathop{\rm K}%
$ (see Fig. 4). At higher bias voltage, $V>\Delta _{Nb}$, there is a
deficient current in the $I-V$ curve, which indicates the tunneling regime
of junction conductivity.\cite{BTK,Zaitsev}

In the BTK model (Ref. \onlinecite{BTK}) the barrier strength is
characterized by a dimensionless parameter $Z=H/\hslash v_{F}$, where $v_{F}$
is the quasiparticle Fermi velocity and $H$ is the interface repulsive
potential, responsible for an interfacial scattering. In our particular case 
$Z$ value is characterized by the ratio of the Fermi velocities between Au
and YBCO, $v_{F}$(Au)$/v_{F}$(YBCO).\cite{Zaitsev} Simulations of the $I-V$
curve made on the basis of the BTK model with fitting parameters $T=4.2%
\mathop{\rm K}%
$ and $\Delta _{Nb}=1.2$m$%
\mathop{\rm V}%
$ for $Z=0,1$ and $5$ (see dashed curves, Fig.3) result in a good agreement
with the experimental ones for $Z\gtrsim 5$ . Since there is a very small
difference between $I-V$-curves for $Z\gtrsim 5$ in the BTK model, one can
correspondingly estimate the lower limit for $Z$ to be $5$ and the average
transparency $D=1/(1+Z^{2})$ to be $4\times 10^{-2}$. The average
transparency can be presented also as $D\sim (2\rho _{c}l)/(3r$),\cite
{Zaitsev,DeGennes} where $\rho _{c}$ - a specific resistivity of the YBCO
film in $c$-direction, $l$ - mean free path along the $c$-axis of YBCO.
Taking into account typical values $\rho _{c}\sim 10^{-2}%
\mathop{\rm %
\Omega}%
\cdot 
\mathop{\rm cm}%
$, $l\sim 1%
\mathop{\rm nm}%
$, $R_{N}\sim 1%
\mathop{\rm %
\Omega}%
$, $S=100\mu $m$^{2}$, one can roughly estimate $D$ to be $10^{-3}$. This
evaluation is in agreement with the comparison of the experimental $I-V$
curves with the BTK model given above.

Actually, a mixed order parameter symmetry with a dominant $d$-wave
component and a significant $s$-wave component had been confirmed.\cite
{Kouznetsov} The YBCO is orthorhombic and a small difference of the
magnitude of the order parameter in the $a$- and $b$-direction is expected.
The twins in $ab$ planes of heavily twinned YBCO can decrease this
difference because of sign changes of the $s$-wave order parameter across
twin boundaries. But because of the disorder in the YBCO surface layer and
reducing of the $d$-wave component across the twin boundary the $d$-wave
component is much more suppressed than the $s$-wave one. The suppression can
give rise to a thickness of a coherent $s$-wave surface layer and non-random
tunneling current.\cite{Haslinger,Golubov98}

Besides, an anisotropy of the pair potential in $s$-wave superconductor (for
Nb see Ref.\onlinecite{mac}) enhances the first harmonic component of CPR.
In Ref. \onlinecite{Rae} Rae proposed that this effect induced the first
harmonic component of CPR in Bi$_{2}$Sr$_{2}$CaCu$_{2}$O$_{8+\delta }$
(BSCCO)-Pb Josephson junctions without assuming $s$-wave component in BSCCO.
Thus, one cannot expect the $I_{1}$ component to be exactly zero.

CPR measurements were performed using a single-junction interferometer
configuration in which a junction of interest is inserted into a
superconducting loop with an inductance $L\approx 80$p$%
\mathop{\rm H}%
$.\cite{Rifkin} By coupling the interferometer to a tank circuit through the
mutual inductance, we measure the impedance of the tank
circuit-interferometer system as a function of the external magnetic flux of
the interferometer $\Phi _{e}$. As shown in Ref. \onlinecite{Ilichev98} the
CPR can be extracted from the following equations: 
\begin{equation}
\varphi =\varphi _{e}-\beta f(\varphi ),  \label{eq:phi}
\end{equation}
\begin{equation}
\tan \alpha \approx \frac{k^{2}Q\beta f\prime (\varphi )}{1+\beta f\prime
(\varphi )},  \label{eq:tan}
\end{equation}
where $\varphi _{e}=2\pi \Phi _{e}/\Phi _{0}$, $\alpha $ is the phase angle
between the drive current $I_{rf}$ and the tank voltage $U$ at the resonant
frequency, $\beta =2\pi LI_{c}/\Phi _{0}$ is a normalized critical current
and $Q$ is the quality factor of the tank circuit. The details of this
technique were reported in Ref. \onlinecite{Ilichev99b}. Note, that from Eqs.%
\ref{eq:phi} and \ref{eq:tan} follow: 
\begin{equation}
\tan \alpha \approx k^{2}Q\beta \frac{df(\varphi )}{d\varphi _{e}}.
\label{eq:tand}
\end{equation}
Therefore, the method outlined above is a differential method in respect to
the CPR. This provides a high sensitivity of the CPR measurement. The second
harmonic component of the CPR can be detected with an amplitude down to $100%
\mathop{\rm nA}%
$.\cite{Ilichev00} Thus, for investigated c-axis Nb/Au/YBCO junctions we
estimate that a second harmonic component much lower than the first one can
be detected, $I_{2}/I_{1}>10^{-2}$. A typical CPR dependence measured at $T=5%
\mathop{\rm K}%
$ is presented in Fig.~5. No deviation of the CPR from a sinusoidal $2\pi $
periodic function was found in any of the four samples that were
investigated within the temperature range $1.6%
\mathop{\rm K}%
<T<8.5%
\mathop{\rm K}%
$. It means that the Josephson current is due to first order tunneling which
agrees with microwave experiments.\cite{Kleiner} It should be pointed out,
that our method is more sensitive than in Ref. \onlinecite{Kleiner}.

Quite generally, $I_{2}/I_{1}\sim D$, where $D$ is the transparency of the
tunnel barrier. If the transparency is low (tunneling limit) $I_{2}$ is
negligible compare to $I_{1}$. To be able to measure the second harmonic
component $I_{2}$, a tunnel junction with high transparency is required.

Let us discuss the difference in CPR for $c$-axis Nb/Au/YBCO and asymmetric $%
45^{\circ }$ bicrystal junctions, investigated in Ref. \onlinecite{Ilichev99}%
.

An asymmetric $45^{\circ }$ grain boundary Josephson junction prepared on a
bicrystal substrate satisfies such a demand of high transparency, as
documented in Ref. \onlinecite{Ilichev99}. In that case, a Fourier analysis
showed that $I_{n}$ is non-negligible also for $n\geq 3$ which indicates a
transparency $D$ close to $1$ and the CPR measurements exhibit a robust $\pi 
$ periodic component of $I_{s}(\phi )$.\cite{Ilichev99} The formation of the
Andreev bound states mentioned above even enhances the second harmonic
component $I_{2}$.\cite{Tanaka97}

There are no Andreev bound states for $c$-axis tunnel junction. In the $c$%
-direction of YBCO a short coherence length and a Fermi velocity mismatch
between low-$T_{c}$ and high-$T_{c}$ superconductors result in the
transparency much lower compare to $ab$ plane junction.

From the normal-state interface resistivity ($r$) one can estimate the ratio
between transparencies of the grain boundary ($D_{GB}$) and the $c$-axis ($%
D_{c}$) tunnel junction, $D_{GB}/D_{c}\sim 10^{2}$. Such a significant
difference between transparencies can explain the absence of a second
harmonic component in $c$-axis tunnel junction as the magnitude $I_{2}\sim
DI_{1}$ is below the resolution of the method.

The clear observation of a Josephson current with a pure $\sin (\varphi )$
CPR in a Nb/Au/YBCO tunnel junction can be interpreted as an indication of a
dominant $s$-wave component in the order parameter of bulk YBCO.\cite{Sun}
The reason for such a conclusion, that often is given in the literature, is
that for a pure $d$-wave order parameter, the symmetry of the $c$-axis
tunneling between $s$-wave and $d$-wave superconductors leads to $I_{1}=0$. 
\cite{Tanaka94} On the other hand, the $I_{2}$ term results from
higher-order tunneling processes and one can neglect its weak angular
dependence, i.e. $I_{2}\neq 0$. It is present also in weak links based on
conventional $s$-wave superconductors, but in the tunnel limit $%
|I_{2}/I_{1}|\ll 1$. However, the CPR is $2\pi $ periodic and we can exclude
that current is due to higher-order tunneling.

To conclude, we have found the first order tunneling of the Josephson
current of Nb/Au/YBCO tunnel junction due to a non-negligible $s$-wave
component in YBCO. We have shown experimentally, that the second harmonic
component $I_{2}$ corresponding to the second order tunneling, if present,
is smaller than $1\%$ of $I_{1}$. The absence of the second harmonic
component of CPR can be caused by a low transparency of the $c$-axis
Nb/Au/YBCO tunnel junction resulting from Fermi velocity mismatch between Nb
and YBCO in the $c$-direction of YBCO and by the suppression of the $d$-wave
component of the order parameter in the surface layer of YBCO.\cite
{Haslinger,Golubov98,Rae}

We would like to thank P. Dmitriev for Nb films deposition, K. I.
Constantinyan and M. Q. Huang for assistance in measurements, A. Tzalenchuk,
F. Lombardi and P. Mozhaev for fruitful discussions and prof. Tord Claeson
for critical reading of the manuscript. This work was supported by INTAS
program of EU (Grant No11459), the DFG (Ho461/3-1), the Swedish Material
Consortium of superconductivity, the Russian Foundation of Fundamental
Research and the Russian National Program on Modern Problems of Condensed
Matter. Partial support by $D$-wave Systems, Inc. is gratefully acknowledged.

\end{document}